# Multi-channel target speech enhancement based on ERB-scaled spatial coherence features

Yicheng Hsu[1]; Yonghan Lee[2]; Mingsian R. Bai[3]

[1] Department of Power Mechanical Engineering, National Tsing Hua University, Taiwan

[2] Department of Power Mechanical Engineering, National Tsing Hua University, Taiwan

[3] Electrical Engineering, National Tsing Hua University, Taiwan

**ABSTRACT**

Recently, speech enhancement technologies that are based on deep learning have received considerable research attention. If the spatial information in microphone signals is exploited, microphone arrays can be advantageous under some adverse acoustic conditions compared with single-microphone systems. However, multichannel speech enhancement is often performed in the short-time Fourier transform (STFT) domain, which renders the enhancement approach computationally expensive. To remedy this problem, we propose a novel equivalent rectangular bandwidth (ERB)-scaled spatial coherence feature that is dependent on the target speaker activity between two ERB bands. Experiments conducted using a four-microphone array in a reverberant environment, which involved speech interference, demonstrated the efficacy of the proposed system. This study also demonstrated that a network that was trained with the ERB-scaled spatial feature was robust against variations in the geometry and number of the microphones in the array.

## 1. INTRODUCTION

Recently, monaural speech enhancement technologies that are based on deep learning have demonstrated promising results compared with traditional signal processing methods. Most advanced approaches operate in the short-time Fourier transform (STFT) domain, and they estimate real-valued masks [1, 2] or complex masks [3] using a deep neural network. However, time–frequency-masking-based approaches are ineffective in canceling out noise between speech harmonics, and they consume considerable computational resource. To solve these problems, PercepNet [4] uses a triangular equivalent rectangular bandwidth (ERB) filter band, and it applies a comb filter for the finer enhancement of periodic components of speech. DeepFilterNet [5], which is a two-stage speech enhancement framework, uses ERB-scaled gains to enhance the speech envelope, and it employs deep filtering to enhance periodic components. Moreover, target speech enhancement methods [6–8] that utilize auxiliary speaker information to address privacy issues and handle overlapped speech conditions are increasingly being developed. The auxiliary information can be obtained from pre-enrolled utterances of the target speaker [9–11], video imagery of the target speaker [12], electroencephalogram signals [13], and speech activities of the target speaker [14].

Although the aforementioned monaural approaches are effective in extracting close-talking speech, the signal received at the microphone is heavily distorted in far-field applications such as hands-free teleconferencing and smart speakers. Speech enhancement performance can be improved by the use of additional spatial information, which can be provided by a microphone array. For instance, the direction-aware SpeakerBeam presented in [15] combines an attention mechanism with beamforming. The neural spatial filter proposed in [16] uses directional information while extracting the target speech. The time-domain SpeakerBeam (TD-SpeakerBeam) presented in [17] employs interchannel phase differences as additional input features to increase the speaker separation capability. Instead of *ad hoc* spatial features [18, 19], suitable spatial feature estimates can be obtained from multichannel microphone signals by using a trainable spatial encoder. Although these methods can utilize spatial information, the training models can be used only for a microphone array that is identical to the training set. Therefore, a previous study introduced an array-geometry-agnostic personal speech enhancement model [20] that works regardless of the number of microphones that are used or the type of array configuration that is applied. Moreover, we recently proposed a target speech sifting network [21] that is based on a long-short-term spatial coherence (LSTSC) feature; target speaker enrollment data were used to demonstrated the effectiveness of the network in speech enhancement, and it remains robust when changes are made to the array geometry and number of microphones used. However, to the best of our knowledge, no study has investigated ERB-scaled spatial information.

---

[1] shane.ychsu@gapp.nthu.edu.tw
[2] yong-han@gapp.nthu.edu.tw
[3] msbai@pme.nthu.edu.tw

To address the aforementioned research gap, the present study developed a geometry-agnostic multichannel target speech enhancement system, which utilizes spatial features and pre-enrolled utterances as inputs, based on DeepFilterNet [5]. In this system, a novel ERB-scaled LSTSC feature is computed as a spatial feature in relation to the speaker activity pertaining to each ERB band, which is derived from our previous work [21]. Spatially varying source signals are extracted by the ERB encoder layers, and speaker-dependent information is used to further extract the target speaker signals from the mixture. To assess the effectiveness of the proposed system, we explored its robustness against unseen array geometries and determined the influence of the number of microphones, including a single-microphone setup with no spatial information, on its performance. Short-time objective intelligibility (STOI) [22] and perceptual evaluation of speech quality (PESQ) [23] were employed as performance metrics in the experiments.

## 2. MULTI-CHANNEL TARGET SPEECH ENHANCEMENT SYSTEM

### 2.1 LSTSC feature

Consider one static interference source and one target speaker in a reverberant room. The signals are received by a microphone array containing $M$ elements and are analyzed in the STFT domain. Assume that the target speaker and a spatially stationary interference source coexist in a room. The problem can be formulated in the STFT domain, with $l$ denoting the time index and $f$ denoting the frequency index. The signal captured by the $m$th microphone can be expressed as

$$Y^m(l,f) = \sum_{j=1}^{J} A_j^m(f) S_j(l,f) + V^m(l,f), \quad (1)$$

where $m \in \{1,...,M\}$ denotes a given microphone, $l \in \{1,...,T\}$ denotes the time frame, $f \in \{1,...,F\}$ denotes the frequency bin, $Y_j^m(l,f) = A_j^m(f) S_j(l,f)$ denotes the signal of the $j$th source measured by the $m$th microphone, $A_j^m(f)$ denotes the acoustic transfer function relating the $j$th source and the $m$th microphone, $S_j(l,f)$ denotes the signal of the $j$th source, and $V^m(l,f)$ denotes the nondirectional noise measured by the $m$th microphone.

For each TF bin, the short-term relative transfer function (RTF) between the $m$th microphone and reference microphone 1 can be estimated by averaging ($R + 1$) frames:

$$\tilde{R}^m(l,f) \equiv \frac{\hat{\Phi}_{y^m y^1}}{\hat{\Phi}_{y^1 y^1}} \equiv \frac{\sum_{n=l-R/2}^{l+R/2} Y^m(n,f) Y^{1*}(n,f)}{\sum_{n=l-R/2}^{l+R/2} Y^1(n,f) Y^{1*}(n,f)}, \quad (2)$$

where * denotes the complex conjugate operation, $\hat{\Phi}_{y^m y^1}$ denotes the short-time cross-spectral density estimate between channels $m$ and 1, and $\hat{\Phi}_{y^1 y^1}$ denotes the short-time autospectral density of the reference microphone. A "whitened" feature vector $\mathbf{r}(l,f) \in \mathbb{R}^{M-1}$ pertaining to each TF bin can be calculated as follows:

$$\mathbf{r}(l,f) = \left[ \frac{\tilde{R}^2(l,f)}{|\tilde{R}^2(l,f)|}, ..., \frac{\tilde{R}^M(l,f)}{|\tilde{R}^M(l,f)|} \right]^T, \quad (3)$$

where $|\cdot|$ is the complex modulus.

For a spatially stationary interference source, the following long-term RTF (which is computed through recursive averaging) can be used to approximate the expectation of time-average of the feature vector:

$$\bar{r}^m(l,f) = \lambda \bar{r}^m(l-1,f) + (1-\lambda)r^m(l,f), \quad m = 2,\ldots,M, \tag{4}$$

where $\lambda$ is the forgetting factor that facilitates the tuning between the global view (large $\lambda$) and the local view (small $\lambda$) of time frames. The feature vector $\bar{\mathbf{r}}(l,f)$ is also whitened after each recursive step:

$$\bar{\mathbf{r}}(l,f) = \left[\frac{\bar{r}^2(l,f)}{|\bar{r}^2(l,f)|}, \ldots, \frac{\bar{r}^M(l,f)}{|\bar{r}^M(l,f)|}\right]^T. \tag{5}$$

To fully exploit the temporal–spatial information conveyed by the whitened RTF, [21] we can calculate the LSTSC, $\gamma_{lf}(l,f)$, between the short-term whitened feature vector $\mathbf{r}(l,f)$ and the long-term whitened feature vector $\bar{\mathbf{r}}(l,f)$ as follows:

$$\gamma(l,f) \approx \frac{1}{M-1}\sum_{m=2}^{M}\frac{\mathrm{Re}\{\tilde{R}^m(l,f)\bar{r}^m(l,f)^*\}}{|\tilde{R}^m(l,f)||\bar{r}^m(l,f)|} \approx \frac{1}{M-1}\mathrm{Re}\{\mathbf{r}^H(l,f)\bar{\mathbf{r}}(l,f)\}, \tag{6}$$

where $\mathrm{Re}\{\cdot\}$ denotes the real-part operator. The Euclidean angle [24] is adopted in the LSTSC definition to ensure sign sensitivity. The LSTSC is an indicator of the spatial correlation between the short-term RTF and long-term RTF, which are associated with the spatially stationary interference. LSTSC values with a large $\lambda$ (global LSTSC) are used to sift out TF bins that either correspond to the active target or correspond to both the target and the interference, which are rendered inactive. By contrast, LSTSC values with a small $\lambda$ (local LSTSC) are used to identify TF bins that correspond to the directional sources.

## 2.2 ERB-scaled LSTSC

In this section, we present the novel ERB-scaled LSTSC feature derived from the aforementioned LSTSC feature. On the basis of the equivalent rectangular bandwidth (ERB) for human hearing [25], the dimensions of a noisy signal can be reduced to 16 bands:

$$Y_{ERB}(l,b) = \sum_{f}^{F_b} w_b(f)|Y(l,f)|^2, \quad b \in [0,16], \tag{7}$$

where $w_b(f)$ is the weight of the frequency bins for band $b$ and $F_b$ is the number of frequency bins for band $b$. Therefore, the dimensions of the LSTSC feature can also be reduced to 16 bands of an ERB-scaled spatial coherence feature:

$$\gamma_{ERB}(l,b) = \frac{1}{\pi_b}\sum_{f}^{F_b} w_b(f)\gamma(l,f), \quad b \in [0,16], \tag{8}$$

where $\pi_b$ denotes a weight normalization and is expressed as

$$\pi_b = \sum_{f}^{F_b} w_b(f). \tag{9}$$

## 2.3 Speaker encoder

The speaker encoder generates a speaker embedding vector from pre-enrolled utterances of the target speaker. A speaker embedding can be extracted using a speaker encoder that is trained using the target speech model or a pretrained model for the extraction of speaker information such as the i-vector [26], x-vector [27], or d-vector [28]. In this study, we used the d-vector, which has been successfully used in various applications such as speaker diarization, speech synthesis, personal voice activity detection, and source separation. The proposed model, containing a three-layer long–short-term memory network followed by a projection layer, was trained using a generalized end-to-end loss function [28], and the speaker encoder was trained using the VoxCeleb2 data set [29]. The model yields embeddings in sliding windows. The resulting aggregated embedding, which is known as the d-vector, encodes the target speaker's voice characteristics.

### 2.4 ERB-based multi-channel target speech sifting network

DeepFilterNet [5] is a two-stage deep filtering approach that uses a complex filtering instead of a point-wise multiplication with a mask. In this study, DeepFilterNet was extended yield to multichannel personalized DeepFilterNet. As illustrated in Fig. 1, this network has four inputs: the real and imaginary parts of complex spectrogram features, ERB-scaled spectral feature computed for the reference microphone, d-vector of the target speaker generated by the speaker encoder, and ERB-scaled spatial coherence calculated from the array signal. In this network, an ERB encoder/decoder architecture is used to predict ERB-scaled gains, which can suppress the spatially stationary persistent interference. The d-vector is concatenated to the middle layer to sift the latent features pertaining to the target speaker. To further enhance the periodic components, DeepFilterNet predicts per-bin filter coefficients.

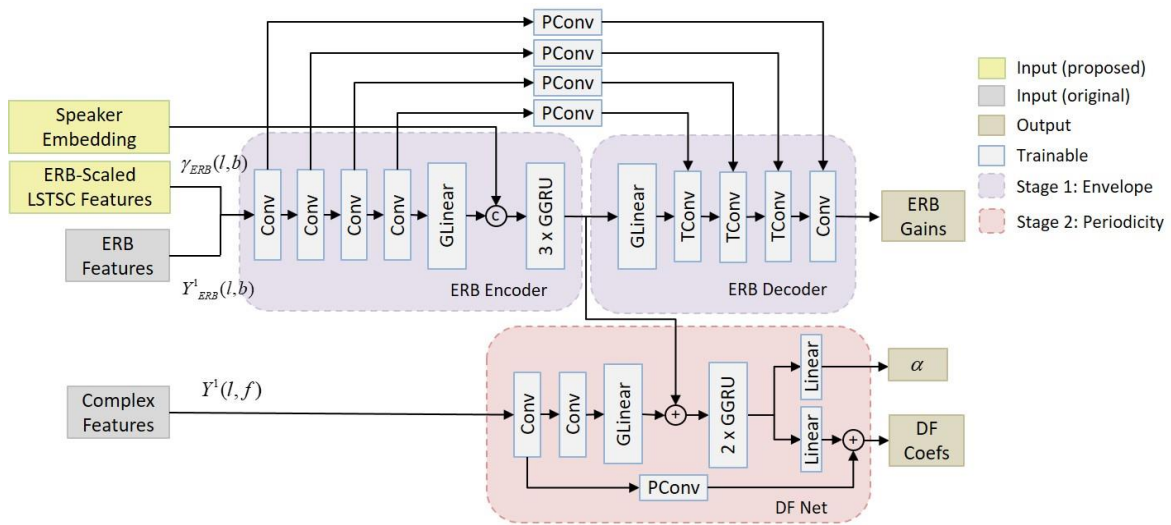

Figure 1 – ERB-based multi-channel target speech sifting architecture

## 3. EXPERIMENTAL STUDY

### 3.1 Data preparation

To validate the proposed multichannel target speech enhancement system, we trained our model using simulated room impulse responses (RIRs) and tested it using measured RIRs. The measured RIRs were based on the recently proposed Tampere University Rotated Circular Array Dataset [30], which contains 777 RIRs. The simulated RIRs were generated using the image-source method [31], with reverberation time (T60) being set to 0.32, 0.48, and 0.60 s. The microphone array was placed at the center of the room, and the target speaker and interference were randomly positioned around the array (within 0.5–2.5 m) under the assumption that the target speaker is always close to the array. In the training and validation sets, a four-element uniform circular array (UCA) of a radius of 4 cm was chosen. In the test set, three array configurations were employed to evaluate the robustness of the proposed enhancement system. The array geometry is illustrated in Fig. 2.

We used data from publicly available data sets and convolved them with the measured and simulated RIRs. Clean utterances that are necessary for training and testing were selected from the train-clean-360 and dev-clean subsets of the LibriSpeech corpus [32], which contain utterances from 921 and 40 speakers, respectively. We generated noisy training and test data using the VoxConverse data set [33], from which 74-h human-conversation clips from YouTube were chosen. The audio contained noise of various types, such as background noise, music, laughter, and applause.

In the training phase, a reference speech signal, which was derived from the utterances of the target speaker and different from the clean signal, was randomly selected for enrollment. Noisy audio signals in the form of 8-s clips were prepared by mixing clean target speech signals and interferences to yield signal-to-noise ratios (SNRs) of −5, 0, 5, and 10 dB. The sampling rate for all utterances was 16 kHz. Furthermore, the sample size of the training, validation, and test sets were 50,000, 5,000, and 5,000, respectively. The STFT settings were a 32-ms window length, a 16-ms hop size, and a 512-point fast-Fourier transform. The 16-dimensional ERB spectral feature was calculated from the noisy signal that was captured by the reference microphonem and the proposed ERB-scaled spatial coherence feature was calculated on the basis of the ERB-scaled LSTSC in Eq. (8). In this experiment, the forgetting factors were set at $\lambda = 0.999$ and $0.01$ to calculate the global ERB-scaled spatial coherence (ERB-G-LSTSC) feature and the local ERB-scaled (ERB-L-LSTSC) feature.

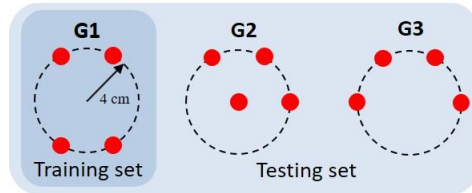

Figure 2 – Array geometries that were utilized in the training and test sets.

### 3.2 Results and discussions

Figure 3 presents the performance of the proposed model for different array geometries at different SNRs. We observed a considerable improvement in the performance of the proposed model when array geometries that were identical to the training set were applied. Furthermore, the proposed model was effective when array geometries that were not used during training were applied. These results suggest that the multichannel target speech enhancement system based on the proposed ERB-scaled LSTSC feature is robust against array geometry variations, which is a desirable property in real-world applications.

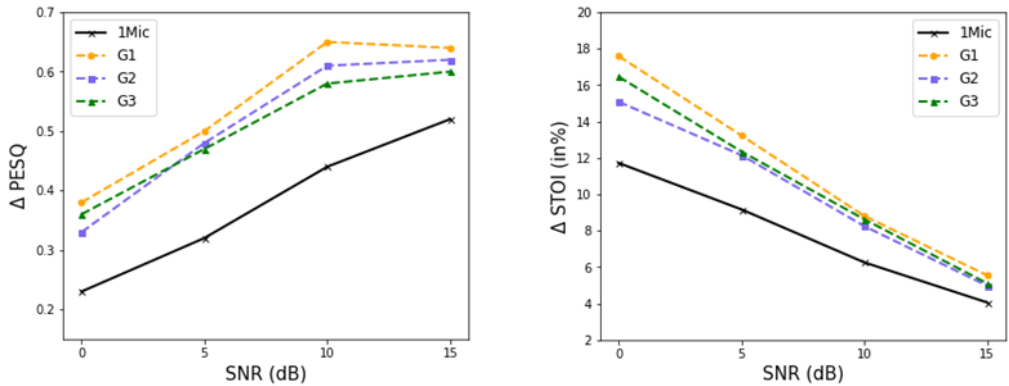

Figure 3 – PESQ and STOI scores for different array geometries.

To determine the influence of the number of microphones on the proposed system, four different configurations with one to four elements (Fig. 4) were used to assess the robustness of the system. For the proposed ERB-scaled spatial coherence feature, a change in the number of microphones engenders a change in only the feature vector dimensions in Eqs. (3) and (5) but not the input dimension in Eq. (8). As indicated in Table 2, increasing the number of microphones was advantageous to enhancement performance at the various SNRs. The monaural speech enhancement system (representing the approach with no spatial information) yielded the lowest performance, especially for scenarios with low SNRs. This result demonstrates that the proposed ERB-scaled LSTSC feature can make the model independent of the array configurations. Without the need to change the model architecture for a specific array configuration, a single model with the proposed ERB-scaled LSTSC feature can be shared by multiple arrays with different array geometries and numbers of microphones.

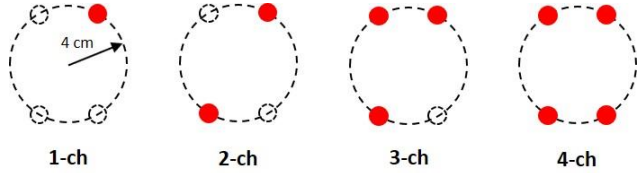

Figure 4 – UCAs with different numbers of microphones.

Table 2 – Comparisons of enhancement performance in terms of PESQ and STOI for different number of microphones.

| Metric | PESQ | | | | STOI (in %) | | | |
|---|---|---|---|---|---|---|---|---|
| SIR (dB) | 0 | 5 | 10 | 15 | 0 | 5 | 10 | 15 |
| Noisy | 1.30 | 1.49 | 1.70 | 2.00 | 61.76 | 72.61 | 81.48 | 87.30 |
| 1-ch | 1.53 | 1.81 | 2.14 | 2.52 | 73.48 | 81.78 | 87.74 | 91.36 |
| 2-ch | 1.64 | 1.91 | 2.22 | 2.51 | 78.60 | 84.73 | 89.09 | 92.00 |
| 3-ch | 1.68 | 1.98 | 2.33 | 2.61 | 78.83 | 85.51 | 89.99 | 92.63 |
| 4-ch | **1.68** | **1.99** | **2.35** | **2.64** | **79.35** | **85.84** | **90.29** | **92.85** |

## 4. CONCLUSIONS

We propose a multichannel target speech enhancement system that is based on ERB-scaled spatial coherence features along with speaker embedding. Our results demonstrate that the novel ERB-scaled spatial feature is useful for target speaker speech enhancement as well as for system robustness against unseen RIRs, unseen array geometries, and changes in the number of microphones used, which is highly desirable for real-world applications. The use of the ERB-scaled LSTSC feature can effectively reduce the computational resource of the proposed speech enhancement system, rendering it compatible with embedded devices.

## ACKNOWLEDGEMENTS

This work was supported by the Ministry of Science and Technology (MOST), Taiwan, under the project number 110-2221-E-007-027-MY3.